\documentclass[usenatbib,onecolumn]{mnras}
\usepackage{graphicx}%
\usepackage{amsmath,amssymb}
\usepackage{comment}

\let\Re\relax
\DeclareMathOperator{\Re}{Re}
\DeclareMathOperator{\atantwo}{arctan_2}
\renewcommand{\vec}[1]{\boldsymbol{\mathrm{#1}}}

\title{Imaging rotating and orbiting exoplanets with the solar gravitational lens}

\author[V. T. Toth and S. G. Turyshev]{
Viktor T. Toth$^1$, Slava G. Turyshev$^{2}$\\
$^1$Ottawa, Ontario K1N 9H5, Canada\\
$^2$Jet Propulsion Laboratory, California Institute of Technology,
4800 Oak Grove Drive, Pasadena, CA 91109-0899, USA}

\date{Accepted XXX. Received YYY; in original form ZZZ}

\pubyear{2023}

\begin{document}

\label{firstpage}
\pagerange{\pageref{firstpage}--\pageref{lastpage}}

\maketitle

\begin{abstract}
We utilize the well-established properties of the solar gravitational lens (SGL) to consider realistic observational scenarios. Actual exoplanets, which may be the target of an SGL observational campaign, are not stationary. Their appearance changes as a result of their diurnal rotation and varying illumination due to their orbital motion around their host star. The nature of the SGL is such that imaging with one telescope is accomplished with a cadence of one pixel at a time, with substantial per-pixel integration times. Therefore, capturing a single snapshot of the target planet with a realistically-sized telescope is not possible. Instead, the planetary surface must be reconstructed by inverting the combined effect of  the SGL's point-spread function and temporal changes induced by the planetary dynamics. Using the Earth as a stand-in, we demonstrate practical feasibility of this approach, by simulating a dynamical system and then recovering topographic images of acceptable quality. The dynamics-induced temporal variability of the exoplanet represents an added challenge, but even in the presence of such dynamics, use of the SGL for exoplanet imaging remains feasible.
\end{abstract}


\section{Introduction}

The solar gravitational lens (SGL) offers an exceptional promise: The ability to produce resolved images of very distant targets, potentially life-bearing exoplanets in distant solar systems \citep{SGL2017}. Yet this promise is predicated on our ability to overcome significant challenges, including imperfections of the lens, substantial
light contamination by the solar corona and other noise sources, not to mention the technical challenge of delivering an autonomous observatory to the SGL focal region where it would have to perform exceptionally accurate navigation in order to maintain position in the context of an SGL-projected image.

Beyond these challenges, which we already addressed in several prior publications \citep{SGL2018,SGL2018a,SGL2018b,SGL2018c,SGL2018d,SGL2018e,SGL2018f,SGL2019a,SGL2019aa,SGL2019b,SGL2020a,SGL2020b,SGL2020c,SGL2020d,SGL2020f,SGL2021a,SGL2021b,SGL2021c,SGL2021d,SGL2021e,SGL2021f,SGL2021g,SGL2021h,SGL2021i,SGL2022a,SGL2022b,SGL2022d,SGL2023a,SGL2023b,SGL2023c}, there remains another, perhaps the toughest of them all: the target's temporal variability. SGL imaging is necessarily done one ``pixel'' at a time, as a suitably equipped spacecraft or constellation of spacecrafts scan a projected image that is several kilometers wide in the focal region of the Sun. The signal-to-noise ratio (determined primarily by a faint Einstein ring viewed through the bright solar corona) demands lengthy integration times each pixel, resulting in timeframes measured in months, if not years, while a sufficient number of pixels are collected. During this time, an exoplanet similar to the Earth will not only perform hundreds of diurnal rotations, it will have also completed one or more orbits around its host star, with the resulting variability of its surface's illumination.

In previous studies, we investigated in exquisite detail the SGL point-spread function (PSF) and the resulting optical properties of the lens including negative spherical aberration (due to the nature of the gravitational monopole) and astigmatism (due to quadrupole and higher-order corrections to the gravitational monopole). However, we treated the PSF as a stationary map from a source plane (which contains a static image of the target exoplanet) to the image plane (located at the SGL focal region). In the present study we explore a different mapping: the time-dependent map of the exoplanet's actual spherical surface to the image plane of the SGL.

We accomplish this by combining the lens PSF (a characteristic of the lens itself) with the dynamics of the source system. We can do so by recognizing that at any given moment in time, the combination of the lens and the current position and phase of the exoplanet results in a linear mapping of its source pixels onto the image plane in the SGL focal region. However blurry the result is, it nonetheless represents nothing more than a linear combination of the exoplanet surface. This implies that after $N$ observations are made in the image plane at different locations, the surface of the exoplanet can be reconstructed in the form of $M\le N$ surface pixels. A number smaller than $N$ may be chosen to obtain a stable solution, treating the total number of observations as an overdetermined system of equations. This may also help to reduce numerical instabilities due to noise.

Before we can address the magnitude of that noise and its inevitable impact on image reconstruction, we first have to understand how an image of a ``moving target'' can be reconstructed in the first place. This is not trivial, even if the orbital and diurnal motion of the exoplanet are known precisely, which is what we assume. In this paper, we present a systematic approach demonstrating that useful results are achievable in principle and that even a modest number of successful observations can yield rudimentary maps of the exoplanet surface.

This paper is organized as follows. In Section~\ref{sec:challenge} we review the basics of SGL imaging, starting with the basic geometry of the target-Sun-observer configuration. We discuss the monopole lens as well as the extended lens represented by mass multipole moments. We briefly summarize the challenges related to light contamination and noise, and summarize our findings related to the dynamical nature of the configuration that arises as a result of the combined motion of target and lens.

In Section~\ref{sec:model} we take the next step by considering the intrinsic changes in the target's appearance, as a result of its diurnal rotation and orbital motion, which implies varying illumination. Using the Earth as a stand-in, we develop a set of semi-analytical equations that can be used to model these effects, and convolve these equations with the PSF of the SGL itself. We then present and discuss actual results, which demonstrate the basic feasibility of the approach by showing that after a modest number of observations, recognizable images of the target can be recovered at an acceptable signal to noise ratio.

We summarize our findings in Section~\ref{sec:discussion}, where we also briefly mention the remaining challenges that will need to be addressed before using the SGL can become reality.

In the Appendix, we offer a brief description of the software that was used to generate our results.

\section{The SGL challenge}
\label{sec:challenge}

The SGL concept presents us with a remarkable opportunity: detailed imaging of distant worlds at resolutions that are not achievable by any other means using presently available or foreseeable technology.

Although the raw capabilities of the SGL are impressive (light amplification of ${\cal O}(10^9)$, angular resolution of ${\cal O}(0.1)$~nanoarcseconds), there are very substantial obstacles that make observing through the SGL challenging. These include:
\begin{itemize}
\item The distance to the SGL focal region is enormous \citep{SGL2017}. Parallel rays of light grazing the solar disk and bent by the gravitational field of the Sun meet along the focal line at a distance of $\sim$~548 astronomical units (AU). Practical imaging requires observations from an even greater distance from the Sun, to allow a telescope with limited resolving power to distinguish (and block) the solar disk from light coming from the distant source;
\item The image produced by a monopole gravitational lens is ``blurry'', exhibiting significant negative spherical aberration \citep{SGL2019b,SGL2020a,SGL2020b,SGL2020c,SGL2020f};
\item Light from a distant world is mixed with much brighter light from contaminating light sources, including the Sun itself, the solar corona and the host star of the distant source \citep{SGL2018c,SGL2018e,SGL2018f};
\item Additionally, even small deviations from the monopole, such as the Sun's minuscule quadrupole mass moment, can substantially alter the pattern in which light is distributed in the focal region \citep{SGL2021b,SGL2021c,SGL2021d,SGL2021e,SGL2023a};
\item The image that is projected by the Sun is large, typically several kilometers in size, and must be scanned one pixel at a time with the observing spacecraft traversing the image region \citep{SGL2022a,SGL2022b};
\item The projected image is moving noninertially, pivoting as a result of the combined motion of the source and the Sun itself \citep{SGL2021i};
\item The source that is being imaged is not static; its appearance changes over time as a result of its rotation, orbital motion around its host star and changing surface features.
\end{itemize}
We have addressed all these challenges in our efforts to date, except for the last one, which is the subject of the present paper.

But first, we present a summary of our results to date, establishing our methods.

\subsection{The SGL imaging geometry}

As it is well known, the Sun deflects rays of light that pass in the vicinity of the solar disk \citep{Einstein1911,Einstein1915}. The magnitude of this deflection is determined by the equations of general relativity as
$\theta = 2r_g/b$,
where $r_g=2GM_\odot/c^2$ is the Schwarzschild-radius of the Sun, $G$ is Newton's constant of gravitation, $M_\odot$ is the mass of the Sun, $c$ is the vacuum sped of light and $b$ is the light ray's impact parameter. This important prediction was first confirmed by the historic May 1919 solar eclipse expedition of Eddington \citep{Eddington1920}, confirming a deflection of $2r_g/R_\odot\simeq$1.75", twice the value that one would predict using a Newtonian corpuscular theory of light, for light rays originating from a distant star that appears adjacent to the solar disk when it is blocked by the Moon.

Knowledge of this angle leads to the easy calculation of the distance at which light rays with the same impact parameter meet along the focal line:
\begin{align}
z=\frac{b}{\theta}
=\frac{b^2}{2r_g} \simeq 547.8 \Big(\frac{b}{R_\oplus}\Big)^2~{\rm AU}.
\label{eq:z}
\end{align}

As this expression implies, the focal region of the SGL is characterized by a half-line, extending from $\sim 547.8$~AU from the Sun to interstellar distances. Solving (\ref{eq:z}) for $b$ yields the impact parameter associated with a given distance from the Sun:
\begin{align}
b=
\sqrt{2r_gz}.
\end{align}
Given a distant, compact source, a telescope situated on the side of the Sun opposite from that source, at a distance $z>548$~AU, will see light from that source appear as a faint ring (the Einstein ring) around the Sun.

For this ring to be resolvable as distinct from the Sun itself, its angular separation from the Sun must be consistent with the angular resolution of the telescope. That is to say, the following inequality must be satisfied:
\begin{align}
\frac{b-R_\odot}{z}=
\frac{\sqrt{2r_g z}-R_\odot}{z}
>\frac{\lambda}{2d},
\end{align}
where $\lambda$ is the observing wavelength and $d$ is the observing telescope's aperture. Using $\lambda=1~\mu$m and $d=1$~m as typical values, this implies $z\gtrsim 625$~AU. In practice, $z>650$~AU is considered a minimum observing distance, though greater values, $z>750$~AU would be preferred.

These distances are challenging. Up to five times greater than the distance covered by our most distant spacecraft to date, Voyager 1, it might take several centuries for a probe powered by conventional propulsion to reach this region. However, more advanced propulsion concepts exist and are on the very of reaching practical maturity, including small satellites equipped with solar sails, following trajectories with very small perihelia, to take maximum advantage of solar radiation. These probes can accelerate to solar system escape velocities of 20--25~AU/year, making it possible to reach the SGL focal region in 30 years or less \citep{SGL2020e,SGL2022c,SGL2023-SPIE}.

\subsection{The monopole gravitational lens}
\label{subsec:mono}

A monopole gravitational lens suffers from negative spherical aberration. Instead of focusing light from a distant source to a point, light is appears on a focal half line. Any image projected by such a lens will be blurry. The extent to of this blur is characterized in principle by the lens's point-spread function or PSF.

In the general case, modeling the gravitational lens as a thin lens, the refracted electromagnetic wave can be represented accurately by a hypergeometric function. However, a more practical representation exists in the vicinity of the optical axis, which is the region of interest for us insofar as imaging with the SGL is concerned. Within this region, the PSF of the SGL can be well represented by the approximation,
\begin{align}
{\tt PSF}(\rho)=\frac{4\pi^2 r_g}{\lambda}J_0^2\bigg(2\pi\frac{\rho}{\lambda}\sqrt{\frac{2r_g}{z}}\bigg),
\end{align}
where $J_0$ is the zeroeth Bessel function of the first kind and $\rho$ is the distance from the optical axis in the image region.

The form of this PSF is notable for two important reasons. First, we observe the magnitude of the coefficient of $\rho$ (measured in meters) in the argument to the Bessel function: $J_0(2\pi\rho\lambda^{-1}\sqrt{2r_g/z})\simeq J_0(48.9\,\rho)$
 for typical parameter values including $\lambda=10^{-6}$~m and $z\simeq 650$~AU. Because of the properties of $J_0(x)$, this implies a spatial frequency characterizing a pattern of constructive and destructive interference that is measured in mere centimeters in the image region. Second, the characteristic amplitude of this PSF falls off very slowly with distance, proportional to $1/\rho$. This differs markedly from the PSF of a convex lens with no aberration, which would follow a $1/\rho^3$ pattern.

The SGL will be viewed by a meter-class telescope, an aperture much larger than the spatial pattern of the SGL PSF. The brightness of the Einstein ring seen by such a telescope, therefore, will be determined by the average brightness of the centimeter-scale pattern of light and dark regions. It stands to reason that the observed signal, then, may be representable by a suitably averaged form of the SGL PSF. This is indeed the case: We found this averaged PSF in the form of an elliptic integral representation, namely
\begin{align}
{\overline{\tt PSF}}(\rho)=\frac{2}{\pi}~\Re{\tt E}\left(\frac{\tfrac{1}{2}d}{\rho};\frac{\rho}{\tfrac{1}{2}d}\right).
\label{eq:PSF}
\end{align}
where ${\tt E}(z;k)=\int_0^z\sqrt{1-k^2t^2}/\sqrt{1-t^2}~dt$ is the incomplete elliptic integral in the Legendre normal form.

\subsection{The extended lens}
\label{subsec:multi}

The SGL is not a perfect monopole. Its deviations from the monopole are tiny: the dimensionless quadrupole mass moment of the Sun is only $J_2\simeq 2.25\times 10^{-7}$. Yet it cannot be neglected, as even this small contribution adds up as light signals travel the substantial distance from the Sun to the focal region. The displacement of light rays as a result of $J_2$ can be comparable to the size of the projected image of the exoplanet being studied, significantly ``scrambling'' light in the image region.

The PSF of an extended gravitational lens is given by ${\rm PSF}(\vec{x})=|B(\vec{x})|^2$ \citep{SGL2021a,SGL2021b,SGL2021c,SGL2021d,SGL2021e}, with the complex amplitude of the electromagnetic field in the lens plane characterized by
\begin{align}
B({\vec x})&{}=\frac{1}{2\pi}\int_0^{2\pi} d\phi_\xi \exp\bigg[-i\frac{2\pi}{\lambda}\Big(\sqrt{\frac{2r_g}{z}}\rho\cos(\phi_\xi-\phi)+2r_g\sum_{n=2}^\infty \frac{J_n}{n}\left(\frac{R_\odot }{\sqrt{2r_g z}}\right)^n\sin^n\beta_s\cos[n(\phi_\xi-\phi_s)]\Big)\bigg],
\end{align}
where $\vec{x}=(\rho,\phi)$ is a location in the image plane, $J_n$ is a dimensionless zonal harmonic coefficient characterising the $n$-th mass moment of the Sun, while $\phi_s$ and $\beta_s$ are the heliocentric azimuth and colatitude of the observing location.

Instead of focusing light to a blurry central region, this PSF spreads light out in the form of a caustic, dominated by the famous ``astroid'' pattern that is determined by the quadrupole moment. Evaluating this PSF can be computationally costly. Restricting ourselves to the quadrupole-only case, however, we find that with the exception of the exact boundary of the astroid caustic, the PSF can be computed accurately everywhere else using the algebraic solution of a quartic equation, which was found using the method of stationary phase to approximate the integral. The deep connection between the physics of the astroid caustic and the algebra of the solution is demonstrated by the fact that the discriminant of the quartic actually replicates the size and shape of the caustic region. This was an unexpected and deep connection between geometry and algebra, related to the physics of the lens \citep{SGL2021d}.

An imaging telescope that is located within the projected astroid caustic, looking back at the lens, sees the famous ``Einstein cross'' pattern that is characteristic of the quadrupole lens. As the telescope approaches the caustic boundary, the Einstein cross collapses into two spots of light. Outside the caustic, the pattern rapidly drops back to the familiar monopole pattern. Therefore, when the astroid caustic is small, smaller in particular than the anticipated or modeled resolution of the lens, the monopole approximation remains useful, albeit the signal strength may be reduced and correspondingly, the
SNR suffers. On the other hand, imaging at a resolution that is finer than the caustic pattern's characteristic size does not appear feasible. In this case, the light signal in the image region is just too scrambled for reconstruction.

\subsection{Light contamination and noise}

As an optical system, the SGL is inherently noisy. Observations with the SGL amount to measuring the brightness of a faint Einstein ring or Einstein cross that is seen across the bright solar corona, with additional light leakage from the nearby Sun. Moreover, the $1/\rho$ ``blurry'' behavior of the PSF implies also that there might be substantial light contamination from ``interlopers'', bright sources that appear in the sky near the intended target, including the target exoplanet's host star itself.

It is possible to subtract much of this light contamination if simultaneous observations are made from multiple, nearby vantage points. One possible strategy is to use differenced signals for image reconstruction (that is, the arithmetic differences of Einstein ring or Einstein cross brightness measurements made by multiple observing satellites at the same time in the same image plane), which means light contamination from the solar corona (which will appear the same for nearby satellites) will be mostly eliminated and light contamination from interlopers will be reduced. What is unavoidable, however, is shot noise, which is substantial due to the limited number of photons being captured. Shot noise can be reduced by increasing integration times (it is proportional to the inverse square root of integration time) but there, we run into another limitation: extended integration times will produce motion blur due to the dynamics of the observed system and the temporal behavior of the target exoplanet. Other strategies exist, of course, including, for instance, oversampling the image plane and use suitable optimization techniques to reduce noise.

One of the reasons why shot noise is of substantial concern is that reconstruction of a high-resolution image of the target require deconvolution, and the deconvolution of the SGL PSF carries a substantial ``deconvolution penalty'' in the form of a reduction in the SNR. For the monopole lens, we found a reliable estimate of this penalty in the form
\begin{align}
\frac{{\tt SNR}_{\tt R}}{{\tt SNR}_{\tt C}}\sim 0.891\frac{D}{d\sqrt{N}}.
\label{eq:penalty2}
\end{align}
where ${\tt SNR}_{\tt C}$ and ${\tt SNR}_{\tt R}$ represent the signal-to-noise ratio of the convolved and recovered images, respectively, $D$ is the average spacing between pixels as sampled in the image region, and $N$ is the total number of pixels. This estimate was derived from the basic numerical properties of the monopole PSF, and also confirmed through several numerical simulations.

The presence of the quadrupole moment scales this deconvolution penalty in the solar equatorial plane as
\begin{align}
\frac{{\tt SNR}^{\tt quad}_{\tt R}}{{\tt SNR}^{\phantom{\tt quad}}_{\tt C}}\simeq 0.315\frac{\sqrt{2r_g z}}{J_2R_\odot^2}\frac{D}{\sqrt{N}}\simeq\left(\frac{D}{646~{\rm m}}\right)\left(\frac{z}{650~{\rm AU}}\right)^{1/2}\frac{1}{\sqrt{N}}.
\end{align}
At higher solar latitudes, the SNR improves approximately as $1/\sin^2\beta_s$ (where $\beta_s$ is the solar colatitude) until it begins to approach the SNR of the monopole lens as the quadrupole contribution to the penalty becomes negligible.

Numerical simulations confirm the approximate validity of this expression so long as the pixel spacing is substantially larger than the cusp-to-cusp size of the projected astroid caustic in the imaging region.

\subsection{Imaging with the SGL}

As a lens, the Sun projects an image, blurred though it might appear. Given the planetary radius of $R_{\tt p}$, the basic geometry of this image is easy to compute, from the distance from lens to source ($\bar{z}$) and lens to image region ($z$), with image radius given by
\begin{align}
R_{\tt image}=R_{\tt p}\frac{z}{\bar{z}}\simeq 2\left(\frac{R_{\tt p}}{R_\oplus}\right)\left(\frac{z}{650~{\rm AU}}\right)\left(\frac{10~{\rm pc}}{\bar{z}}\right)~{\rm km}.
\end{align}
An image of this side obviously cannot be obtained all at once, as it would require an image sensor, an imaginary ``projection screen'' that is several (or several ten) square kilometers in size. Rather, observing spacecraft would need to scan this image region, measuring the apparent brightness of the Einstein ring that is seen around the Sun at a series of locations.

This creates a unique set of requirements that an observing mission must satisfy. Observations taken hours, days, weeks apart need to be correlated and properly calibrated to ensure that they can be used in the reconstruction of the source. The dynamics and temporal behavior of the source cannot be ignored. Last but not least, the observing spacecraft need to be navigated at high accuracy, as their position (and especially, any changes in their position) relative to the projected image of the exoplanet must be known with precision.

In fact, an imaging mission must begin with finding the projected image of the exoplanet in the first place. Given the distance of the observing spacecraft from terrestrial stations, conventional navigation is of little help. Furthermore, considering the $\sim$1 week round-trip signal travel time, any navigation must be performed autonomously. The spacecraft may have access to astrometric measurements and may have a view of the host star (indeed, for most sensible trajectories, the host star is visible throughout the entire cruise to the focal region) but it must then rely on precise measurements of the brightening of the lensed images of first the host star and then, eventually as it emerges, the target exoplanet in order to find the part of the focal region where the exoplanet image appears.

\subsection{The dynamics of an imaging mission}

Finding the exoplanet image is a challenge on its own right but it is then compounded by the need for precise stationkeeping or, at the very least, precise estimation of the observing spacecrafts' position with respect to the projected exoplanet image.

The reason for this difficulty is the pivoting, or reflex motion of the projected image as a result of the combined motion of the exoplanet and the Sun (i.e., the lens). The exoplanet follows its own orbit around the two-body barycenter of the planet and its host star, perturbed slightly by other bodies in that exosolar system. The observing spacecraft will follow basically unperturbed hyperbolic escape trajectories with respect to the barycenter of the inner solar system. Meanwhile, the Sun (i.e., the lens) moves substantially due to the combined gravitational pull of the planets, in particular Jupiter and the inner planets.

Consequently, the projected image of the exoplanet in the imaging region also moves noninertially. It is not strictly necessary for an observing spacecraft to remain motionless with respect to that image (so long as integration times are kept sufficiently short to avoid motion blur -- note that observations can still be combined later computationally to reduce the impact of stochastic noise). However, the precise position of the observing spacecraft with respect to the projected image must be known accurately at all times in order for observations to be of any value.

This implies keeping track of velocity changes up to several ten m/s and worst-case accelerations of ${\cal O}(10^{-5})$~m/s$^2$. This autonomous navigational accuracy may be achievable using a combination of in-situ observations (possibly using multiple spacecraft), the use of reserve spacecraft to maintain a local inertial reference frame, and refined, fitted orbital models of the exosolar system.

\section{Changing view of an exoplanet}
\label{sec:model}

The problems discussed in the preceding section are well understood and have been analyzed in detail. However, in all the analysis that we performed to date, we quietly assumed that the imaged object is a fully illuminated exoplanet with a static appearance.

This, of course, is a gross oversimplification. In reality, the view of an exoplanet changes over time as a result of the planet's diurnal rotation, as well as its motion around its host star and the resulting changes in illumination.

\begin{figure}
\begin{center}
\includegraphics[scale=1]{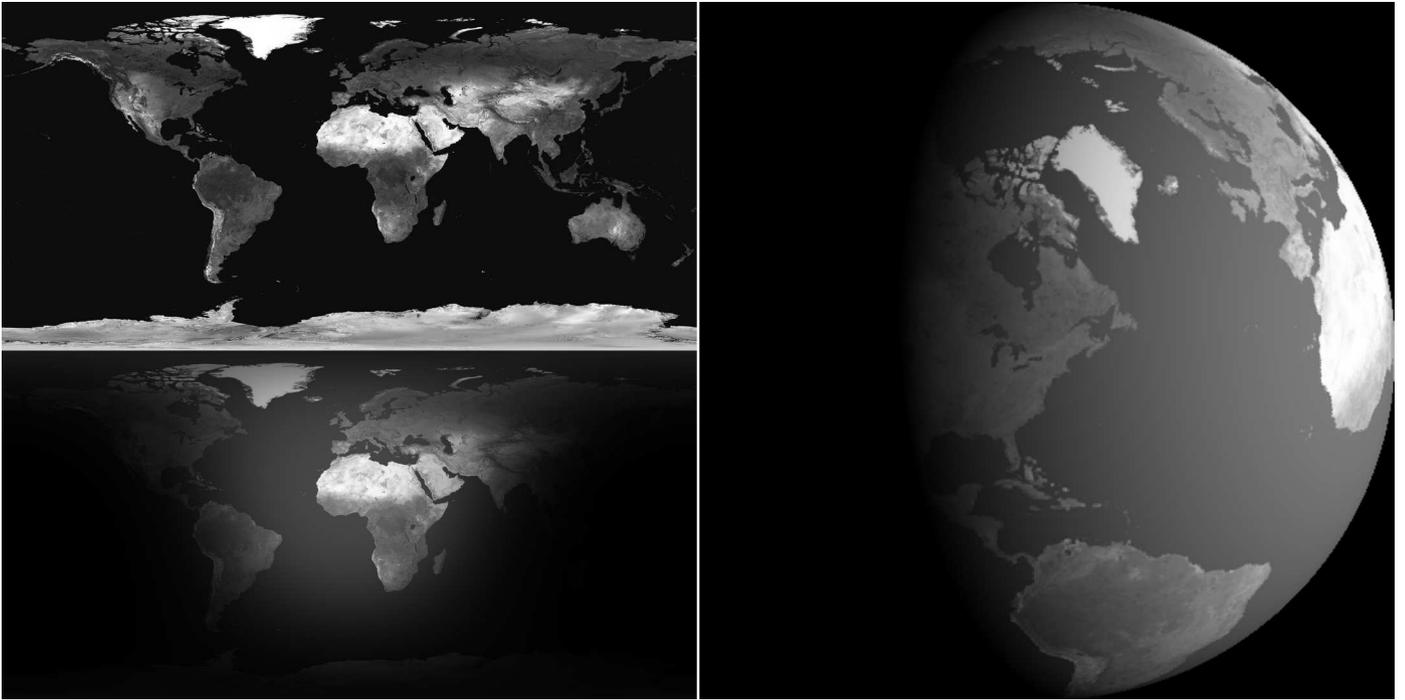}
\end{center}
\caption{\label{fig:combined}Views of the Earth. Top left: Cylindrical projection of the fully illuminated surface of the Earth, with any cloud cover removed. Bottom left: The same, illuminated by the Sun at a specific epoch, assuming a fully Lambertian reflective surface. Right: the illuminated globe, viewed from a vantage point that corresponds to a latitude of 45$^\circ$~N.}
\end{figure}

This is depicted in Figure~\ref{fig:combined}. In this figure and in our analysis, we use the Earth as a stand-in for the exoplanet. The upper left of Fig.~\ref{fig:combined} shows the fully illuminated map of the Earth's surface. Our goal is to recover this view as accurately as possible using a finite number of observations of the Einstein ring that the Earth would project through an SGL-like gravitational lens, viewed by a telescope in that lens's focal region.

The concept of ``rotational deconvolution'' has been addressed in previous studies \citep{SGL2018a,ELF2018}. The emphasis in those cases, however, was different. The objective was either
\begin{itemize}
\item Reconstruction of an albedo map of the full Earth, ``stitching together'' images taken at various times while accounting for differences in illumination;
\item Reconstruction of low-resolution views of exoplanets using a time series of unresolved images.
\end{itemize}

Similar concepts were also explored in the context of ``super-resolution'' imaging: the use of temporal variations in the target's appearance to extract information beyond the optical resolution of the observing instrument \citep{Puschmann2005,Bauer2011,Oberti2022}.

In contrast, our goal is to reconstruct an albedo map of the target planet by using single-pixel observations (corresponding to the position of the observing spacecraft within the image area) that are collected over the course of several days, perhaps weeks or months. Our objective is not to go beyond the capabilities of the SGL but to ensure that the SGL can reliably deliver resolved images of the exoplanet's surface despite the fact that the exoplanet acts as a ``moving target''.

To accomplish our goal, first we must assess how the surface is illuminated at any given time. The lower left of Fig.~\ref{fig:combined} shows the same cylindrical projection as it would actually appear under actual solar illumination. (Note that this is still an idealized case, as in the present analysis, we ignore cloud cover or deviations of the surface from an ideal, diffuse Lambertian reflector.)

This partially illuminated surface is, of course, not flat but rather, the surface of a sphere. The observer's vantage point has a fixed latitude with respect to this sphere, but the observer's longitude changes as the sphere (the planet) rotates. Figure~\ref{fig:combined} right shows the resulting view. This is what a sufficiently powerful telescope would see if it had the resolving power. Unfortunately, such a telescope would require an aperture measured in the tens of thousands of kilometers, clearly an impossibility.

Instead, we have the SGL, which has the aperture, but which is also blurry. The resulting view, given in Figure~\ref{fig:blurred}, is what can actually be observed, one pixel at a time. Meanwhile, the planet rotates and its illumination changes.

Our goal, therefore, is to model all these effects, combining them into an appropriate mapping of the planetary surface into observational data. We begin with modeling the illumination of the planet as a function of location on the planetary surface and time. For this and subsequent calculations, we use well-developed formulas that are specific to the Earth; it goes without saying that similar expressions (modified for convenience if the dynamics of the exoplanetary system differ markedly from that of our solar system) must be developed for the actual observing target.

\subsection{Modeling the illumination}
\label{subsec:illum}

To compute illumination at a given location terrestrial at a given time, we need to know the solar elevation at that location at that time. Well-known, widely used semianalytical formulas are available for this purpose\footnote{See \url{https://en.wikipedia.org/wiki/Position_of_the_Sun} for further details.}. In our example, we begin by calculating the mean solar longitude:
\begin{align}
L=I+v\cdot n = 280.460^\circ + 0.9856474^\circ\cdot n,
\end{align}
where $I=280.460^\circ$ is the mean solar longitude at the J2000 epoch (January 1, 2000, 12:00 UTC, $J=2451545.0$ Julian date), and $v=0.9856474^\circ$ is the mean daily motion of the Sun along the ecliptic. The value of $n$ is the number of days since the J2000 epoch.

From this, we calculate the mean solar anomaly:
\begin{align}
g = g_0+v_g\cdot n=357.528^\circ + 0.9856003^\circ \cdot n
\end{align}
where $g_0=357.528^\circ$ is the mean solar anomaly at the J2000 epoch, while $v_g=0.9856003^\circ$ is the mean daily motion of the Sun in its elliptical orbit.

We then compute the ecliptic longitude:
\begin{align}
\lambda_e = L + \lambda_1 \sin(g) + \lambda_2 \cdot \sin(2g)= L + 1.915^\circ \cdot \sin(g) + 0.020^\circ \cdot \sin(2g),
\end{align}
where $\lambda_1=1.915^\circ$ is the amplitude of the largest term in the equation of the center, which accounts for the ellipticity of the Earth's orbit around the Sun, and $\lambda_2=0.020^\circ$ is the amplitude of the second-largest term in the equation of the center, which accounts for the smaller perturbations in the Earth's orbit.

We next calculate the obliquity of the ecliptic:
\begin{align}
\epsilon = \epsilon_0+\delta\epsilon \cdot n=23.439^\circ - 0.0000004^\circ  \cdot n,
\end{align}
where $\epsilon_0=23.439^\circ$ is the mean obliquity of the ecliptic at the J2000 epoch, whereas $\delta\epsilon=-0.0000004^\circ$ is the rate of change per day of the obliquity of the ecliptic.

We can now calculate the right ascension, $\alpha$, and declination, $\delta$, of the Sun:
\begin{align}
\alpha &{} = \arctan\left(\frac{\cos(\epsilon) \cdot \sin(\lambda_e)}{\cos(\lambda_e)}\right),\\
\delta &{} = \arcsin\Big(\sin(\epsilon) \cdot \sin(\lambda_e)\Big).
\end{align}
The local sidereal time is calculated as
\begin{align}
T_{\rm LS} = T_0 + v \cdot n + \text{LON} + 15T_{\rm UT} =  100.46^\circ + 0.9856474 \cdot n + \text{LON} + 15 \cdot T_{\rm UT},
\end{align}
where $T_0=100.46^\circ$ This is the Greenwich Mean Sidereal Time at the J2000 epoch, measured in degrees. This yields the hour angle
\begin{align}
H = T_{\rm LS} - \alpha.
\end{align}

Finally, we can calculate the azimuth ($A$) and altitude ($h$) of the Sun:
\begin{align}
A &{} = \arctan\left(\frac{\sin(H)}{\cos(H) \cdot \sin(\text{LAT}) - \tan(\delta) \cdot \cos(\text{LAT})}\right),\\
h &{} = \arcsin\Big(\sin(\text{LAT}) \cdot \sin(\delta) + \cos(\text{LAT}) \cdot \cos(\delta) \cdot \cos(H)\Big).
\end{align}

In particular, $\sin{h}$ yields the illumination of any surface element when $h$ is positive. If $h$ is negative, that means that the Sun is below the horizon at that location and the illumination is zero. The map at the bottom right of Fig.~\ref{fig:combined} and similar maps were produced using this formulation.

Of course this formulation is specific to the case of the Earth. For an exoplanet, a suitable coordinate system will need to be established and a similar formulation will need to be created, the details of which depend on the dynamics of the exosolar system. While details may differ, we expect the principles to remain largely the same as in the case of the Earth.

\subsection{Modeling the globe}
\label{subsec:globe}

Once we have a surface map of the planet with illumination, it is possible to construct a view of the planetary globe as seen from a given direction. As seen from a rotating exoplanet, our Sun will always be at the same latitude $\phi_0$ in a geographic coordinate system associated with that planet. The longitude $\lambda$, however, will change in sync with the planet's diurnal rotation.

Therefore, the point directly underneath the Sun on that exoplanet will have the coordinates
\begin{align}
\phi {}& = \phi_0,\\
\lambda {}& = \lambda_0 - \delta\lambda,
\end{align}
where $\delta\lambda$ is the current phase of the planet relative to the epoch, i.e., the fractional sidereal day. (The negative sign, in the case of the Earth, is a result of the convention that the longitude that corresponds to a fixed celestial position decreases over the course of the day as the Earth spins eastward.)

To establish a view of this planet from the Sun's direction, we can then use standard formulas for great circle navigation. Considering a view of the planetary globe and a Cartesian coordinate system $(x,y)$ with its origin at the center of the view of the globe, the surface coordinates $(\phi_2, \lambda_2)$ on the planetary surface that correspond to $(x,y)$ in the image plane can be calculated as\footnote{See \url{https://en.wikipedia.org/wiki/Atan2} for the definition of $\atantwo(y,x).$}
{}
\begin{align}
\gamma &{} = \arcsin\frac{\sqrt{x^2+y^2}}{R_\oplus},\\
\theta &{} = \atantwo(y,x),\\
\phi_2 &{} = \arcsin\big(\sin(\phi) \cos(\gamma) + \cos(\phi) \sin(\gamma) \cos(\theta)\big),\\
\lambda_2 &{} = \lambda + \atantwo\big(\sin(\theta) \sin(\gamma) \cos(\phi), \cos(\gamma) - \sin(\phi) \sin(\phi_2)\big).
\end{align}

This formulation can be used to map the illuminated cylindrical projection, as shown in Fig.~\ref{fig:combined} lower left, into the appropriately illuminated, rotated view of the globe that is shown in Fig.~\ref{fig:combined} right.

\subsection{Constructing an effective PSF}

In the preceding subsections, we established the geometry of the exoplanet's illuminated surface (using our own Earth as a stand-in) and its projection to a view of its globe as seen by a distant telescope.

\begin{figure}
\begin{center}
\includegraphics[scale=1.5]{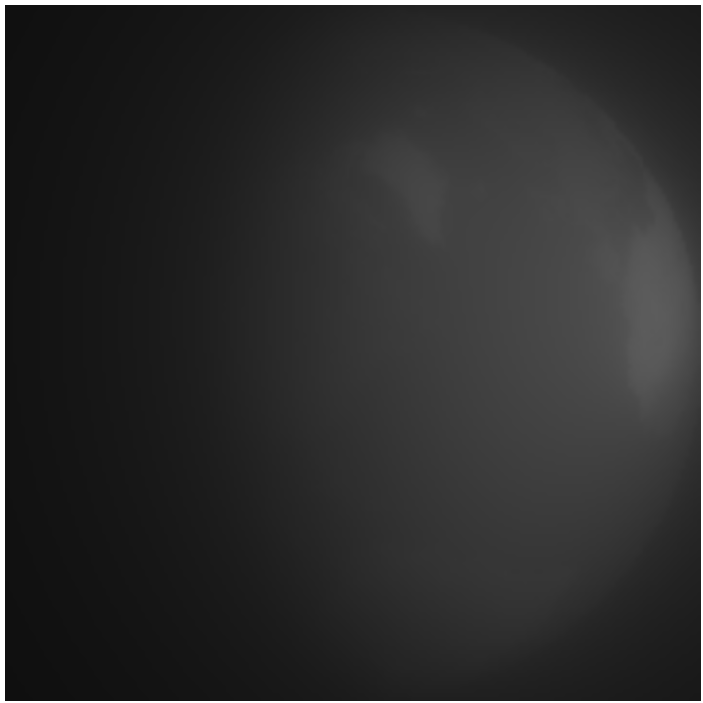}
\end{center}
\caption{\label{fig:blurred}The actual projected view of an Earth-like planet by a gravitational lens. Rather than the sharp image that a ``perfect'' telescope might show, the image is heavily blurred by the spherically aberrated, astigmatic lens.}
\end{figure}

However, the image shown in Fig.~\ref{fig:combined} right is not the kind of image that the blurry SGL projects. The actual SGL image is more akin to that shown in Fig.~\ref{fig:blurred}, as it is blurred by the spherically aberrated, astigmatic gravitational lens.

What this amounts to is that the geometric mapping from surface to (unblurred) image must then be convolved with the mapping from unblurred source to blurred projection in the image region, due to the SGL PSF.

This creates what amounts to an ``effective PSF'', a hybrid mapping: the geometric part is dependent on time and the dynamics of the source, whereas the SGL part is a static property of the gravitational lens itself. Nonetheless, this effective mapping works: At any given time, when our observing satellite samples a pixel in the SGL image region, the combination of the source system geometry and the SGL PSF tells us exactly how much each surface element of the planet contributed to the measured light intensity at that location at that time.

This mapping is strictly linear. The total amount of light measured at any given moment is a linear combination of the brightness of all the surface elements of the exoplanet. Our expectation is that this mapping is, in fact, invertible, and that a brightness map of the source planet at a given resolution can be reconstructed once we are in possession of a sufficient number of measurements of the Einstein ring luminosity as seen from the image plane, precisely stamped with the time of observation and location relative to the projected image.

This narrative can be expressed more formally as follows:

\begin{enumerate}
\item The observing station assumes position $\vec{x}'(t)$ in the image plane at time $t$.
\item A measurement, $I'(\vec{x}'(t))$ is made of the intensity of the Einstein ring as seen at that time from that location.
\item The measurement can be expressed as the convolution of the SGL PSF, ${\tt PSF}(\vec{x}'(t),\vec{x}_0)$, which is a static property of the SGL with no explicit time dependence, and the brightness in the source plane $I_0(\vec{x}_0,t)$:
    \begin{align}
    I(\vec{x}'(t))=\iint~d^2\vec{x}_0~{\tt PSF}(\vec{x}'(t),\vec{x}_0)I_0(\vec{x}_0,t).
    \end{align}
\item The position $\vec{x}_0$ corresponds to a position $\vec{x}$ on the planetary surface with a time-dependent mapping between the two: $\vec{x}_0=\vec{x}_0(\vec{x},t)$.
\item The measurement can thereafter be re-expressed using positions on the planetary surface by a change of integration variables:
    \begin{align}
    I(\vec{x}'(t))=\iint~d^2\vec{x}~{\tt PSF}(\vec{x}'(t),\vec{x}_0(\vec{x},t))I(\vec{x},t),
    \label{eq:II}
    \end{align}
    where $I(\vec{x},t)$ now represents the apparent brightness of the planetary surface at $\vec{x}$ at time $t$, as seen from the viewing direction. However, we note that if the surface is Lambertian, $I(\vec{x},t)$ has no dependence on the viewing direction, which effectively means that the change of integration variables in Eq.~(\ref{eq:II}) introduced no additional contribution.
\item Finally, $I(\vec{x},t)$ can be expressed as $S(\vec{x})R(\vec{x},t)$ where $S(\vec{x})$ is the intrinsic brightness of the source at location $\vec{x}$, while $R(\vec{x},t)$ is the time-dependent irradiance or illumination of that surface element by the planet's host star.
\end{enumerate}
The form of Eq.~(\ref{eq:II}) makes it clear that the mapping between $I(\vec{x},t)$ and $I'(\vec{x}'(t))$ is linear. Discretizing the integral in the form of a summation over $M$ source pixels on the planetary surface, then, we obtain the expression
\begin{align}
I'(\vec{x}'(t_k))=\sum_{i=1}^M C(\vec{x}_i,\vec{x}'(t_k))S(\vec{x}_i),
\end{align}
or simply
\begin{align}
I'_k=\sum_{i=1}^M C^i_kS_i,
\label{eq:tosolve}
\end{align}
where the factors $C^i_k=C(\vec{x}_i,\vec{x}'(t_k))$, aka. the convolution matrix, are given by the expression
\begin{align}
C^i_k = {\tt PSF}\big(\vec{x}'(t_k),\vec{x}_0(\vec{x}_i,t_k)\big)R(\vec{x}_i,t_k).
\label{eq:convolve}
\end{align}

The PSF of the SGL is known, as discussed in subsections \ref{subsec:mono} and \ref{subsec:multi}. The factor $R(\vec{x},t)$ can be calculated as shown in subsection \ref{subsec:illum}. The mapping from planetary surface to image plane, $\vec{x}_0(\vec{x},t)$ is provided in subsection \ref{subsec:globe}. Therefore, the convolution matrix is known.

Consequently, if we make $M$ observations in the image plane such that $1\le k \le N \le M$, Eq.~(\ref{eq:convolve}) can be inverted (or pseudoinverted if it is an overdetermined system with $N<M$) and the source brightness map $S_i$ can be obtained.

Equation (\ref{eq:convolve}) makes it explicitly clear that all time dependence is now captured in the convolution matrix $C_i^k$. The values $S_i$ represent the unchanging topographic view of the planetary surface. Our goal is to reconstruct $S_i$ from the $I_k'$ observations, e.g., by computing a pseudo-inverse of $C^i_k$ or by equivalent means. The values of $C^i_k$ are known based on the dynamics of the system.

As the system is discretized, reconstruction necessarily involves noise: quantization noise in combination with the noise associated with the ``resampling'' that occurs as we map, e.g., a cylindrically projected topographic map of the source onto the visible spherical surface of the planet. Furthermore, as some areas of the planet are poorly illuminated at any given time, image reconstruction (which effectively normalizes signal levels) will amplify this quantization noise.

Additionally, there is sampling noise: the projected image is measured at specific locations by a telescope with an aperature that is much likely significantly smaller than the effective pixel spacing in the image plane. What this means is that even after many samples are taken, much of the image plane remains unsampled. How representative are the sampling locations? Inevitably, random samples will pick sampling locations that are intrinsically fainter or brighter than their typical neighborhoods. This, too, introduces noise. Curiously, this problem is mitigated somewhat by the blurring due to the SGL PSF itself so we find ourselves in a situation where the image of a more distant planet (which appears as a smaller, more blurred projection in the image plane) may be more easily recoverable, albeit of course at a lower resolution.

As a consequence of these considerations, it is legitimate to be concerned that image reconstruction may amplify noise to levels that will render any attempt to recover of the planetary surface futile.

Such noise is hard to quantify analytically, This was, in fact, our main motivation to turn towards a computer simulation. Using the Earth as a stand-in and the relationship between surface topography and observations as captured by the convolution matrix and Eq.~(\ref{eq:convolve}), we attempted to invert (\ref{eq:convolve}) and solve for $S_i$.

\subsection{Simulation and results}

How effective is the approach that we described in the previous subsection? This is best answered by putting the method to use, utilizing actual images of the Earth as a stand-in and implementing the dynamical model of planetary rotation and illumination, along with projection by the SGL PSF, in software. We can then sample the resulting images, and check to see if the original map of the Earth can be reconstructed on this basis.

\begin{figure}
\begin{center}
\includegraphics[scale=1.5]{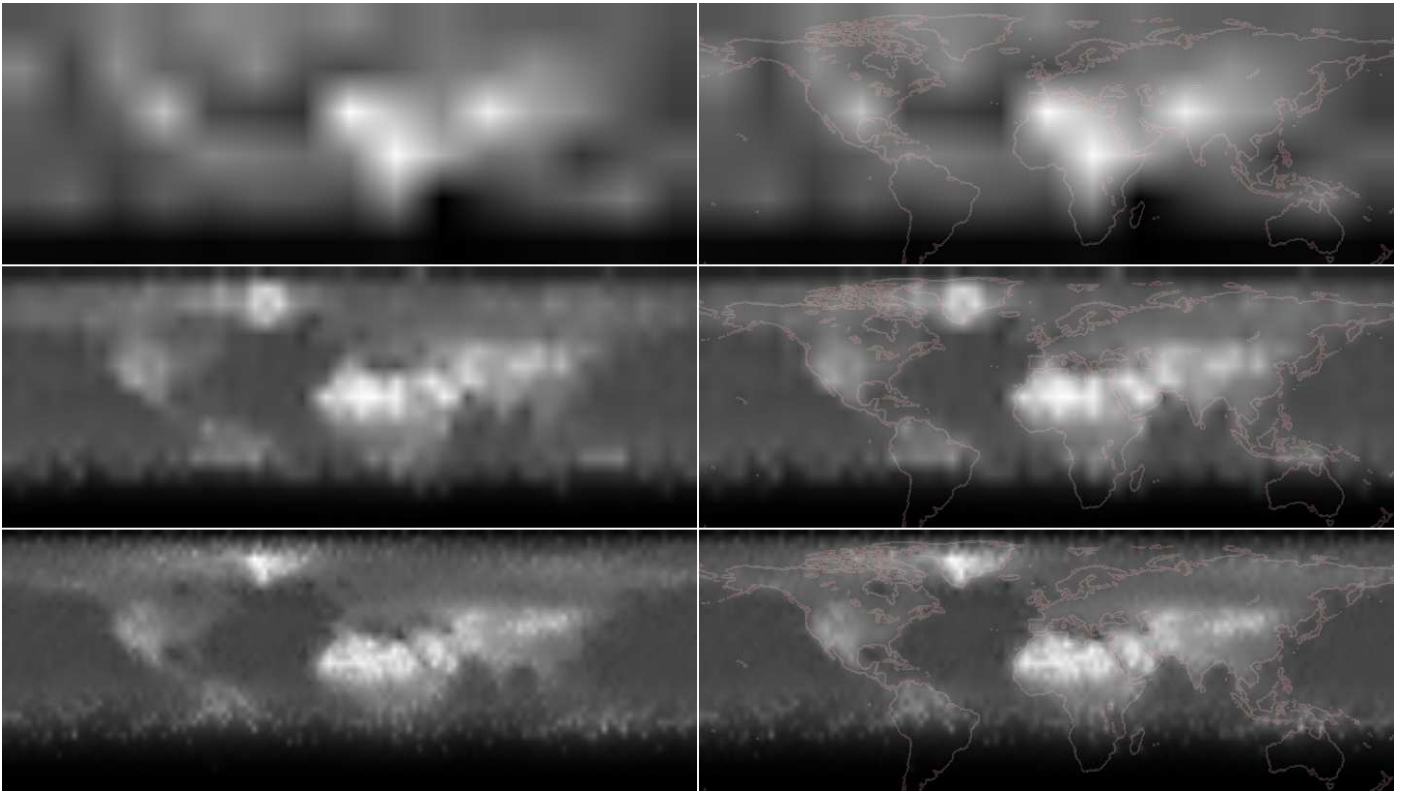}
\end{center}
\caption{\label{fig:resolved3}Progressive reconstruction of the visible Earth surface from simulated observations. The observing location is at 45$^\circ$~N terrestrial latitude, so locations south of 45$^\circ$~S are not visible. Left: the raw reconstructed image. Right: same, with the outlines of the continents shown. Top row: A $6\times 15$ pixel image (smoothed) constructed from 96 observations. Middle row: A $17\times 59$ pixel image from just over 1000 observations. Bottom row: a $40\times 139$ pixel image, from over 5500 observations. In the last image in particular, continental outlines are clearly recognizable along with smaller features, such as the Mediterranean basin, Hudson's Bay, or the Arabian peninsula. Even the northern part of the Australian continent becomes faintly visible, though southern regions remain permanently obscured as the simulation took place during summer in the Northern Hemisphere.}
\end{figure}

This is exactly what we have done, with the results depicted in Fig.~\ref{fig:resolved3}. What this figure shows is the Earth as seen by a distant gravitational lens located at 45$^\circ$~N in the terrestrial reference frame. We allowed the Earth to perform several diurnal rotations, during which we sampled the projected, blurred image at regular simulated intervals, at random locations (basically, the user was clicking the mouse several times in the blurred image to take samples before progressing the image by another simulated hour.)

Once a sufficient number of data points was obtained, we attempted to reconstruct a surface map. To do this, we treated the observations as an overdetermined system, for reasons that we shall explain below. That is to say, if $N$ observations were made, we divided the Earth into $M<N$ pixels, and then solved for the $M$ pixels using a simple optimization algorithm, experimenting with the algorithm's nonphysical parameters such as its solution step size to achieve stable and optimal results.

The results, shown in Fig.~\ref{fig:resolved3}, speak for themselves. After less than 100 simulated observations, a very low resolution image of the Earth was reconstructed. Even at this resolution, the major landmasses are kind of identifiable, though not with high confidence.

Progressing to $\sim$1000 simulated observations, the reconstructed image offers far more detail. North and South America, Eurasia and Africa can be clearly discerned, and even smaller features begin to emerge: Greenland, the Mediterranean basin, the Arabian and the Indian peninsulas are readily identifiable. Meanwhile, we can now also clearly discern the effects of a cylindrical projection and those of illumination. Locations at extreme northern latitudes become more blurred, which is expected, whereas locations at southern latitudes remain invisible, due to the fact that all the simulated observations took place during the Northern hemisphere's summer season.

Moving on to even more, $\sim$5500 observations yields a map of significantly improved resolution, in which even smaller features (e.g., Florida, Japan) can now be seen. At southern latitudes, though faint, Australia also begins to emerge. We can of course not see anything south of 45$^\circ$~S, as those latitudes are permanently hidden from us considering our vantage point at 45$^\circ$~N.

Notably, this reconstruction was possible despite using relatively few data points (in past studies, we routinely investigated imaging with resolutions ranging from $128\times 128=16,384$ to $1024\times 1024\simeq 1.05\times 10^6$ data points).

Moreover, the data points were not arranged in a predetermined grid, but were picked at random while the planetary image was undergoing simulated rotation. Under these conditions, we find it remarkable that even with a limited number of sampled data points, it was possible to reconstruct a recognizable view of the Earth's topography.

Quantifying the relative error, the reconstructed image shows a signal-to-noise ratio of
${\tt SNR}\simeq 1.7.$ A reconstructed image of this quality is more than sufficient to identify major surface features unambiguously.

\section{Discussion}
\label{sec:discussion}

In this study, we investigated for the first time the reconstruction of surface features of an exoplanet, viewed through the solar gravitational lens (SGL), while considering temporal changes of the exoplanet, specifically the consequences of its rotation and varying illumination by its host star.

We found that after a few thousand samples, taken while the planet rotates, reliable reconstruction of planetary features was possible. This result was accomplished under the assumption that the planet's orbital and rotational parameters are known in advance.

Our results demonstrate, at least qualitatively, that a method that combines deconvolution of the SGL PSF with temporal/rotational deconvolution is a feasible approach to image recovery. There are, to be sure, questions that remain unanswered, not addressed in our present study. These include light contamination, unpredictable surface changes, unknown surface optical properties such as specularity, and adverse effects from the quadrupole and higher mass moments of the Sun at higher resolutions. Notwithstanding these challenges, the fact that surface features were recovered relatively easily, and that the recovery procedure demonstrates robust behavior as the number of samples increases, indicates that this is a practical approach to SGL imaging and that the temporal behavior of the target exoplanet represents a manageable problem.
In particular, we demonstrated that:
\begin{itemize}
\item Image recovery is possible even from a surprisingly small ($<100$) number of ``pixel'' obsevations;
\item Image recovery can be progressive: The more data points we collect, the better resolution we obtain (or the more we can improve the SNR);
\item The recovery process is robust, improving the chances of a successful mission even in the face of technical challenges;
\item Sampling locations need not be in a neat, prearranged grid: The probe(s) can crisscross the projected image at convenience, simplifying mission design, reducing fuel consumption, prolonging system lifetime (probe positions still need to be known very accurately relative to the projected image, but they do not need to be precisely controlled).
\end{itemize}

Moreover, in the present study we just used a simple steepest descent optimization algorithm to quickly find an effective solution. A more sophisticated approach may begin, for instance, with a more suitable map projection of the target planet, perhaps a projection that is area-preserving. It may be beneficial to use weighted samples, to characterize the relative contributions of different locations in the projected view of the planetary globe. In addition to solving for the overall reflectivity of source pixels on the exoplanet's surface, these improvements might also help address the remaining challenges listed above.

For now, though, the results shown in Fig.~\ref{fig:resolved3} speak for themselves: a combination of deconvolution of the SGL PSF and time-dependent, ``rotational'' deconvolution that unrolls the combined effect of a planet's diurnal rotation and time-varying illumination, is feasible and yields valuable results.

\section*{Acknowledgements}

The work described here, in part, was carried out at the Jet Propulsion Laboratory, California Institute of Technology, under a contract with the National Aeronautics and Space Administration. VTT acknowledges the generous support of Plamen Vasilev and other Patreon patrons.

\section*{Data Availability}

No data was generated and/or analysed to produce this article.

\bibliographystyle{mnras}
\bibliography{SGLTEMP}

\appendix

\section{Software implementation}

\begin{figure}
\begin{center}
\includegraphics[scale=1.5]{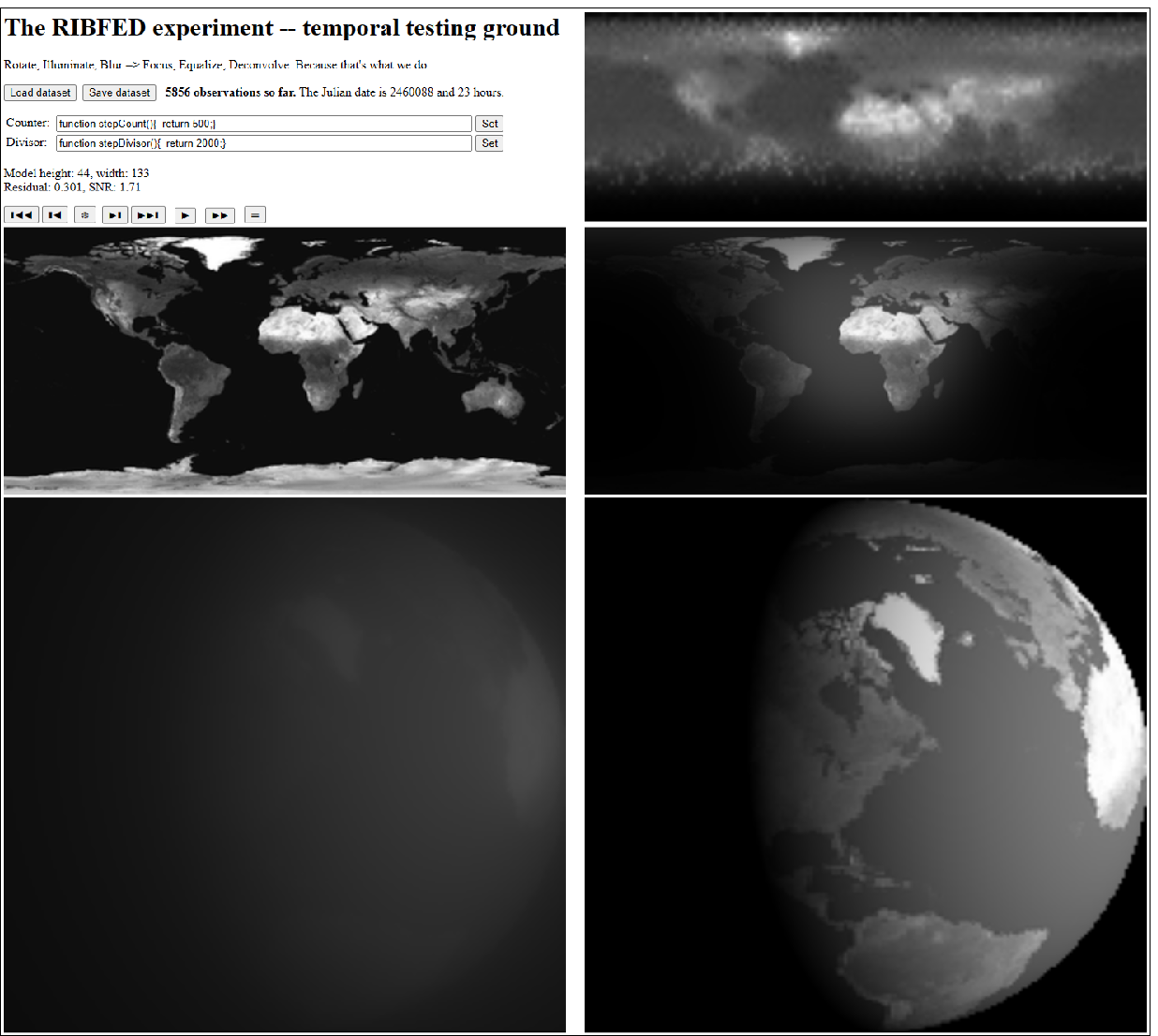}
\end{center}
\caption{\label{fig:sim}Software experiment demonstrating the convolution of varying illumination of an exoplanet as a result of orbital and diurnal dynamics with blurring by the SGL, and successful deconvolution. In this user interface, the top left image is the fully illuminated topography of the planet (using the Earth as a stand-in); to its right, the same but now as illuminated by the Sun on a specific Julian date. Bottom right shows the unobstructed view of the planetary globe from a specific vantage point, whereas the bottom left shows the same, as blurred by the SGL. The reconstructed surface topography appears top right. Various buttons can be used to progress the simulation time, perform convolution with the SGL PSF, or attempt deconvolution.}
\end{figure}

The simulation software that was used to obtain the results presented in this paper was written in JavaScript with a simple HTML front-end, shown in Fig.~\ref{fig:sim}. The software is made available through GitHub: \url{https://github.com/vttoth/TEMPORAL}.

This simulation, which was used to create the images presented in this paper and to obtain the results that support our main conclusions, begins with a monochromatic topographic view of the Earth. The simulation can work at arbitrary resolution, but higher resolution results are computationally costly. We found that using a source topography at a $1^\circ$ resolution is sufficient and the computational costs are modest.

Most of the implementation was done using JavaScript, relying on the fact that modern Web browsers have very efficient JavaScript engines with JIT (just-in-time) compilation capability. Initial image generation is effectively instantaneous, allowing us to even perform animations of the rotating planet with varying illumination at visually pleasing frame rates. However, computing the blur by the SGL is computationally more costly. Given an image of $M\times N$ pixels, a total of $(M\times N)^2$ multiplications by the SGL PSF must be carried out.

Once a blurred image is obtained, the user is invited to click on this image several times to simulate sampling of the image region by observing spacecraft, one pixel at a time. In a typical test run, the simulation is then progressed one hour, the blurred image is recalculated, and the user clicks again several times to obtain additional simulated observations. Once a sufficient number of pixels are sampled, preferably through at least one full rotation of the planet to ensure that its entire visible surface is accounted for, an attempt can be made to obtain a resolved solution. The software automatically calculates an optimal resolution for the reconstructed image, treating the set of observations as a slightly overdetermined system, to guarantee numerical stability of the solution. The solution is a simple steepest descent solver, for which the step size and the maximum number of steps can be changed by the user by defining new functions that compute these values. In most practical cases, however, we found that a step divisor of 2000 and a step count of 500 produce robust, reproducible results.

The source image and the reconstructed image are internally both represented as floating point numbers in the range $[0,1]$. The full Earth topographic map has an average brightness of 0.2. However, the section of the map that corresponds to the visible surface seen by a distant observer whose position corresponds to $45^\circ$~N latitude excludes bright, ice-covered Antarctica, resulting in an average brightness of $\sim 0.16$. Comparing the RMS error of the simulation depicted in Fig.~\ref{fig:sim} to this value yields the estimated SNR of $\sim 1.7$.

\label{lastpage}

\end{document}